\documentclass[fleqn,10pt]{wlscirep}
\usepackage[utf8]{inputenc}
\usepackage[T1]{fontenc}
\usepackage[version=4]{mhchem}
\usepackage{multirow}

\title{Quantum driven proton diffusion in brucite-like minerals under high pressure.}
\author[1]{Sofiane Schaack}
\author[1]{Philippe Depondt\thanks{corresponding author: \texttt{depondt@insp.jussieu.fr}}}
\author[1]{Simon Huppert}
\author[1]{Fabio Finocchi}
\affil[1]{Institut des NanoSciences de Paris (INSP), Sorbonne Universit\'e, CNRS-UMR 7588, 75005 Paris, France}


\begin{abstract}
We investigate the elementary steps at the microscopic level for proton diffusion in brucite under high pressure, which results from a complex interplay between two processes: the O-H reorientations motion around the $\mathbf c$ axis and O-H covalent bond dissociations. First-principle path-integral molecular dynamics simulations reveal that the increasing pressure tends to lock the former motion, while, in contrast, it activates the latter which is mainly triggered by nuclear quantum effects. These two competing effects therefore give rise to a pressure sweet spot for proton diffusion within the mineral. 
In brucite \ce{Mg(OH)2}, proton diffusion reaches a maximum for pressures close to 70GPa, while the structurally similar portlandite \ce{Ca(OH)2} never shows proton diffusion within the pressure range and time scale that we explored. 
We analyze the different behaviors of brucite and portlandite, which might constitute two prototypes for other minerals with the same structure.

\end{abstract}

\begin{document}

\flushbottom
\maketitle

\thispagestyle{empty}

\section*{Introduction}
Hydroxide minerals play an important role in several problems in geology, surface science or for industrial applications. 
Among them, brucite \ce{Mg(OH)2} can be formed at the interface between periclase MgO and water at ambient conditions \cite{review-dissol,Jordan1999,sharma2004situ}.
The trigonal brucite structure consists of alternating layers along the $\mathbf c$ axis that terminate with hydroxyls (figure \ref{fig:Description_Brucite}). 
This structure is common to other hydroxides of divalent metals, such as \ce{Ca(OH)2}, \ce{Ni(OH)2} and \ce{Cd(OH)2}. 
Portlandite \ce{Ca(OH)2} is the main component of cements and concretes, which motivated a large number of investigations about its elastic properties.
Because of their anisotropic structure, brucite isostructural minerals are much more compressible along the $\mathbf c$ axis than in the other two directions, parallel to the stacks.

\ce{Mg(OH)2} can also act as a water vector in subduction zones, through complex processes that take place within the Earth interior\cite{Bodnar2013_review-water-earth,Peacock1999}.
Therefore, the behavior of brucite and brucite-like minerals at very high pressure has been widely investigated. 
X-ray diffraction of \ce{Mg(OH)2} up to 78 GPa showed that the $c/a$ ratio decreases steadily from ambient pressure up to about 25 GPa and then remains almost constant \cite{fei1993static}.
Those results suggest that the properties of brucite at very high pressure, and in particular the nature of the inter-layer bonding, could differ significantly from ambient conditions. 
Moreover, the hydroxyl groups that are mostly parallel to the $\mathbf c$ axis at ambient conditions slant in three equivalent positions as the inter-layer distance shrinks even under moderate pressure \cite{catti1995static,chakoumatos_2013} or when decreasing temperature \cite{chakoumakos1997low}.
Besides reducing the global symmetry from P$\bar{3}m1$ down to P$\bar{3}$, the slanted OH groups can significantly alter several physical properties of brucites. 
Firstly, it could allow for the formation of hydrogen bonds between the layers \cite{parise1994pressure}, which eventually reinforce under further compression and modify the compressibility along $c$ \cite{ulian2019equation}. 
Secondly, when the protons form a non null $\theta$ angle with the $\mathbf c$ axis, they cannot arrange in a static ordered structure. Such proton disorder, which is closely related to \emph{proton frustration} \cite{raugei1999pressure,mookherjee2006high}, has also been invoked as the reason for the pressure-induced hydrogen sublattice amorphization in brucites \cite{catti1995static,nagai2000compression}.
The existence of a quasi two-dimensional proton liquid in those extreme conditions can be conjectured, but the properties of the whole structure, if stable, have so far escaped a precise characterization. In particular, the occurrence of proton hopping is plausible, but whether this process results in a long-range diffusion is a totally open question.

From the theoretical viewpoint, the previous observations call for a dynamical treatment of the proton arrangement within the brucite structure at high pressure.
Moreover, in such conditions nuclear quantum effects (NQE), that is, all the properties that go beyond a purely classical description of ion dynamics\cite{Markland2018}, such as zero-point energy, ion delocalization and tunneling, can have an important impact on the brucite properties. Its light mass makes hydrogen prone to fast diffusion and subject to significant quantum effects even at room temperature.  

In particular, the high O-H stretching frequency implies a non-negligible zero-point energy of about 0.2 eV that is crucial when hopping through local sites which are separated by barriers that classical nuclei cannot overcome simply by thermal fluctuations \cite{Benoit_Nature_1998,Dammak_2012,Bronstein_2014}. Indeed, the seminal paper by Benoit et al.\cite{Benoit_Nature_1998} demonstrated the importance of NQE at room temperature for the high-pressure phase diagram of ice. Of geophysical interest, NQE play a significant role in the hydrogen-bond symmetrization in \ce{AlOOH} under pressure\cite{bronstein_AlOOH,sano_2018}.

Despite the complexity and computational cost of accounting for the quantum nature of light nuclei in simulations, efficient methods have arisen \cite{Berne1986,Marx_Parinello_1996,Tuckerman_1996,Dammak_QTB_2009} and, stimulated by the increasing interest in often paradoxical nuclear quantum effects \cite{Markland2018}, a new field is rapidly growing with potential applications in an eclectic collection of issues including protonic conduction \cite{liang_1993}, hydrogen in biological matter \cite{rossi_2016}, the phase diagram of hydrous minerals that are involved in the water circulation in the Earth's mantle \cite{bronstein_AlOOH,sano_2018}, or hydrogen storage \cite{zhang_2011,wahiduzzaman_2014}.

Proton diffusion in brucite has already been experimentally addressed in several ways, via hydrogen-deuterium interdiffusion \cite{guo_2013}, quasielastic neutron scattering \cite{okuchi_2018} or electrical conductivity measurements\cite{gasc_2011,guo_2014}, for various pressure and temperature conditions and sample preparations. Several diffusion mechanisms, mainly defect-related \cite{gasc_2011,okuchi_2018}, were suggested, however a detailed description of the basic underlying processes at the atomic scale including nuclear quantum effects in an ideal single crystal is still lacking. We therefore show  in this work the elementary mechanism of hydrogen site-to-site hopping in a perfect single crystal in which defects, vacancies, grain boundaries, etc., are absent: the processes thus described are intrinsically due to the crystalline atomic arrangement. While experimental work to date \cite{guo_2013,okuchi_2018,gasc_2011,guo_2014} invokes mainly defect aided diffusion in real systems for which the processes involved are significantly more complex, we focus on the basic mechanism.

Consequently, we address proton hopping occurring in brucite \ce{Mg(OH)2} that appears to be a specific case as compared with portlandite \ce{Ca(OH)2}, which was the object of a recent theoretical study \cite{dupuis2017quantum}, and is also discussed here. 

The reasons for the different behaviors of these two compounds are elucidated.
We employ classical and Path-Integral Molecular Dynamics  \cite{feynman2010quantum, chandlerJCP1981PI, Ceriotti2010_PILE} (PIMD) to disentangle thermal from nuclear quantum effects in a consistent framework, the first, to our knowledge, fully quantum, both for electrons and nuclei, simulation for brucite.  
Among the various proton diffusion mechanisms that were considered in the past, we confirm that the \emph{relay mechanism} \cite{yaroslavstev_1994,dupuis2017quantum} is at play in brucite under high pressure. This two-step process involves proton hops from one oxygen atom to another and rotations of OH groups. We also show in the following that proton hopping is quantum-driven while OH rotation is mostly thermally activated.

\subsection*{Structural changes under pressure}
\begin{figure}[ht!]
\centering
\includegraphics[width=.8\textwidth]{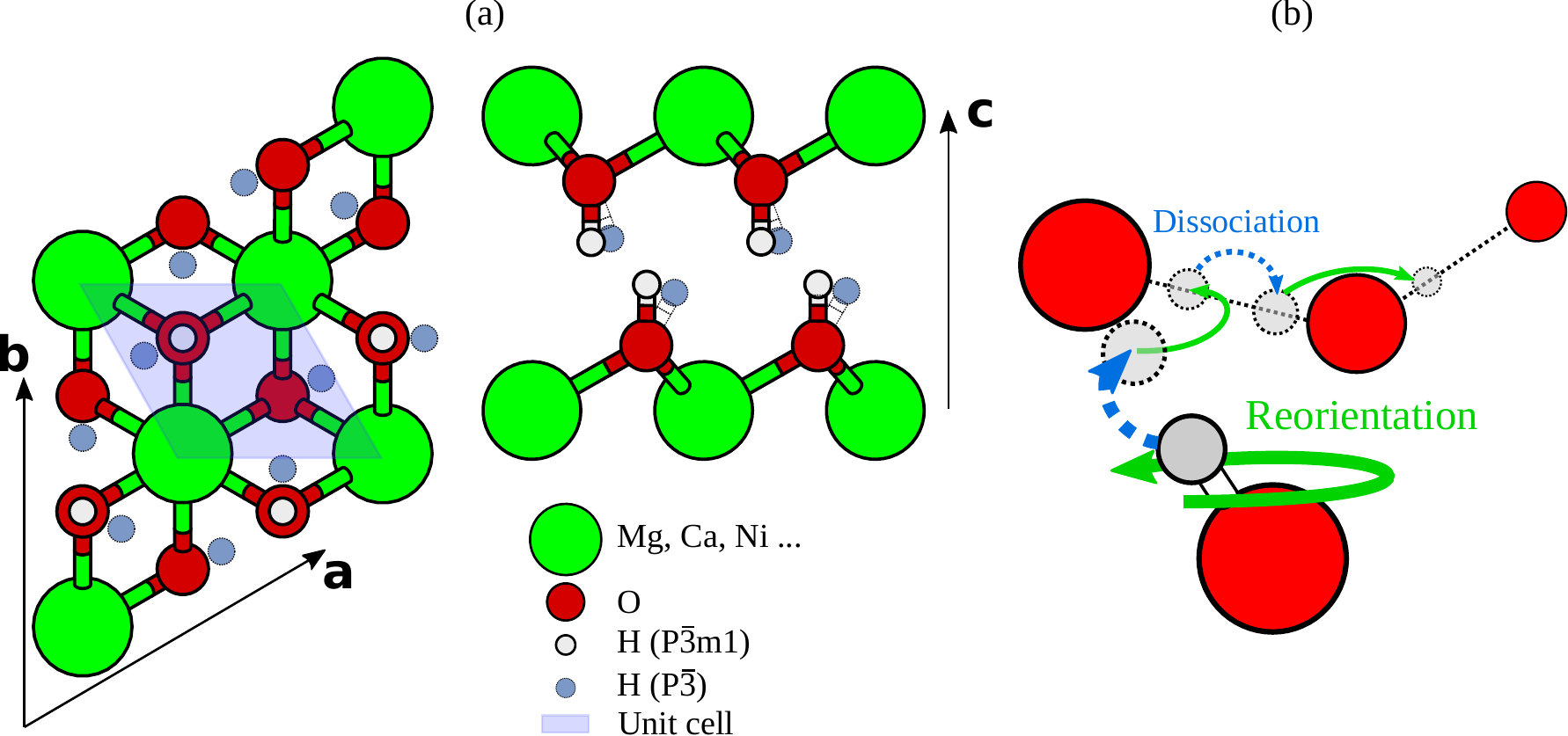}
\caption{Left-hand side: description of brucite minerals structure. The primitive cell is shaded; the simulation box is defined in-plane by vectors $\mathbf{a}=a(3/2,\sqrt{3}/2,0)$ and $\mathbf{b} = a(0,\sqrt{3},0)$, and contains 3 \ce{Mg(OH)2} units. The hydrogen atoms in the P${\bar{3}m1}$ symmetry structure (light grey) are in the $2d$ Wyckoff sites, while the hydrogen atoms in the P$\bar{3}$ structure (dark grey) are in the disordered $6i$ Wyckoff sites. The latter is the stable structure at the pressures of our simulations.
Right-hand side: Proton diffusion mechanism which requires a succession of two alternate steps: proton reorientation (green arrows) i.e. O-H rotation around $c$; and dissociation (blue arrows) i.e. proton hopping between oxygen layers. Oxygen atoms are colored in red, hydrogen atoms in grey. The size of atoms decreases with depth.}
\label{fig:Description_Brucite}
\end{figure}

The brucite minerals are layered structures composed by stacks of metallic ions, oxygen, and hydrogen atoms in a \ce{CaI2}-type structure. \cite{mookherjee2006high,raugei1999pressure,dupuis2017quantum}
The metallic element (\ce{Mg}, \ce{Ca}, \ce{Ni}, \dots) has an impact on several physical parameters, in particular, the lattice parameters and the compressibility of the system are different in the two systems we studied: brucite (\ce{Mg(OH)2}) and portlandite (\ce{Ca(OH)2})\cite{ulian2019equation}.
The metal layers terminate with hydroxyl groups on both sides (figure \ref{fig:Description_Brucite}, left).
At ambient pressure and temperature, the brucite minerals belong to the P$\bar{3}m1$ space group with hydrogen atoms located on the threefold axis above or below oxygen atoms ($2d$ Wyckoff sites) with full occupancy as shown in the left-hand side of figure \ref{fig:Description_Brucite}. 

However, in accordance with previous experimental and simulation results \cite{catti1995static,raugei1999pressure,mookherjee2006high}, as pressure increases, due to the growing repulsive interaction between protons on opposite layers, the protons do not remain above the corresponding oxygen atom.
Brucite minerals adopt a P$\bar{3}$ configuration, where the H nuclei are in the $6i$ Wyckoff sites with a 1/3 occupancy factor (figure \ref{fig:Description_Brucite}, left). More recent X-ray diffraction experiments \cite{chakoumatos_2013} indicate that the P$\bar{3}$ structure with $6i$ sites is plausible even at ambient pressure.
This proton rearrangement has two noticeable consequences:
first, a frustration of the proton orientation upon compression \cite{kruger1989vibrational,duffy1995high,raugei1999pressure}; second, the protons in $6i$ sites could form hydrogen bonds between distinct layers \cite{catti1995static}. This point, which has been raised by several authors, is discussed in the following.

The stability of both brucite and portlandite upon compression has been investigated in the past and a phase transition of portlandite was found through X-ray scattering to occur at approximately 6GPa \cite{iizuka_2011,iizuka_2014}. 
In brucite, however, X-ray experiments\cite{fei1993static} up to 78 GPa at 600 K showed a smooth variation of the diffraction patterns; in the same paper, the authors deduced an equation of state according to which brucite would decompose into periclase and water around 27 GPa at ambient temperature \cite{fei1993static}. These apparently contradictory results suggest a significant kinetic decomposition barrier. 
More recently, a phase transition toward a tetragonal structure was suggested in brucite above 20 GPa by ab initio simulations \cite{hermann2016high}; nonetheless, up to now, no such transition was observed experimentally to our knowledge \cite{duffy1995high,fei1993static}.

Concerning the stability of brucite at pressures beyond 20GPa, there are evidences proving that \ce{Mg(OH)2} could retain its P${\bar{3}}$ structure. First, Johnson and Walker\cite{johnson_1993} showed that brucite is very stable up to P=15GPa and that the slope of the P(T) curve separating brucite from periclase and water is positive. Therefore, a straighforward extrapolation indicates that the dehydration temperature of brucite increases with pressure and should be greater than 600K at any pressure within accessible range, thus much higher than the room temperature where we carried out our simulations. Similarly, other experimental studies\cite{duffy1995high,fei1993static,jiang_2006} suggest that brucite is stable beyond 20GPa, even up to 80GPa and temperatures as high as 600 K. In particular, Jiang et al.\cite{jiang_2006} insisted on the fact that brucite tends to become stiffer and stiffer as pressure increases, with elastic constants becoming close to those of periclase. In table \ref{table:lat_const_brucite}, we report the simulated behavior of the c/a ratio versus pressure in brucite.
\begin{table}[!ht]
\centering\begin{tabular}{|c|cccccccc|}\hline
P [GPa] & 30 & 40 & 50 & 60 & 70 & 80 & 90 & 100 \\ \hline
\rule{0pt}{1em}
$a$ [\AA] & 2.93 & 2.88 & 2.83 & 2.80 & 2.78 & 2.77 & 2.74 & 2.72 \\
$c$ [\AA]& 4.06 & 4.00 & 4.03 & 3.93 & 3.91 & 3.90 & 3.87 & 3.85 \\
$c/a$ & 1.39 & 1.39 & 1.42 & 1.41 & 1.41 & 1.41 & 1.41 & 1.42 \\ \hline
\end{tabular}
\caption{Lattice constants of brucite from our simulations and $c/a$ ratios. The numerical errors are of approximately 0.01\AA\ for $a$, 0.02\AA\ for $c$ and therefore of the order of $10^{-2}$ for $c/a$.} \label{table:lat_const_brucite}
\end{table}
The pressure increase shrinks the distance between the \ce{MgO} layers, due to the important compressibility along the $c$ axis\cite{ulian2019equation}. However, for pressures above 30GPa, the $c/a$ ratio remains approximately constant. This reveals that the compressibility along the $c$ axis becomes comparable to that along $a$\cite{Schaack_JPhysChem_2018}, in connection with the formation of inter-layer hydrogen bonds.
This effect is expected to enhance the cohesion of the system and thus to stabilize the structure, which might be connected to the existence of a rather large barrier hindering brucite decomposition.
Moreover, the formation of inter-layer hydrogen bonds could allow the protons to hop between the two oxygen atoms on the facing \ce{Mg(OH)2} layers. All along the paper, we analyze in detail the reason behind the increasing inter-layer repulsion and discuss the role of the nuclear quantum effects in the formation of hydrogen bonds between the layers.

\subsection*{Proton diffusion mechanism}

As in other hydrous minerals \cite{bronstein_AlOOH,sano_2018}, hydrates \cite{Schaack2019_PNAS} and ices \cite{Benoit_Nature_1998,Bronstein_2016} the proton can hop along hydrogen bonds. 
Following Dupuis et al.\cite{dupuis2017quantum}, we refer to this process here as dissociation, as it implies the breaking of a covalent O-H bond to form another distinct O-H covalent bond. 
However, another process, namely the reorientation (the hopping between the three $6i$ Wyckoff sites in the P$\bar{3}$ space group), is necessary for the proton to move away from the initial O site through the crystal. 
Only the combination of dissociation and reorientation can drive proton diffusion; one of the two processes alone would simply imply a back-and-forth proton motion between neighboring sites.
Proton diffusion in brucite-like minerals is therefore a two-step compound process, intrinsically different from the Grotthuss mechanism that is active in water and other H-bonded hydrates  \cite{grotthuss_1806,yaroslavstev_1994}.

In P$\bar{3}$ \ce{Mg(OH)2}, the proton experiences an effective triple well potential among the $6i$ Wyckoff sites, which controls reorientation within the ($ab$) plane. We refer to these reorientation events as "in-plane" motion.
On the other hand, the "out-of-plane" dissociation mechanism involves an effective double well potential along the O-O direction characterizing the covalent and hydrogen bonds. 
A recent study \cite{dupuis2017quantum} analyzed the proton diffusion mechanism in portlandite for different temperatures and concluded that thermal activation of the reorientation hopping can be the limiting factor for proton diffusion. 
Although the authors guessed that nuclear quantum effects could favor the dissociation mechanism, the latter process was studied only classically.

\section*{Results}

In the following, we mainly focus on the effect of pressure in brucite and unravel the complex and quantum driven proton diffusion in this material, accounting for NQE in a consistent framework for both reorientation and dissociation processes. Rising pressure tends to increase the strength of hydrogen bonds. As a consequence, OH dissociation and proton reorientation display opposite trends with pressure. Furthermore, we show that dissociation is significantly enhanced by NQE while reorientation is mostly thermally activated.
Finally, the behavior of protons in portlandite is analyzed, and differences between Mg(OH)$_2$ and Ca(OH)$_2$ are discussed.

\subsection*{Quasi-2D proton layer}

\begin{figure}[ht!]
\centering
\includegraphics[width=0.7\textwidth]{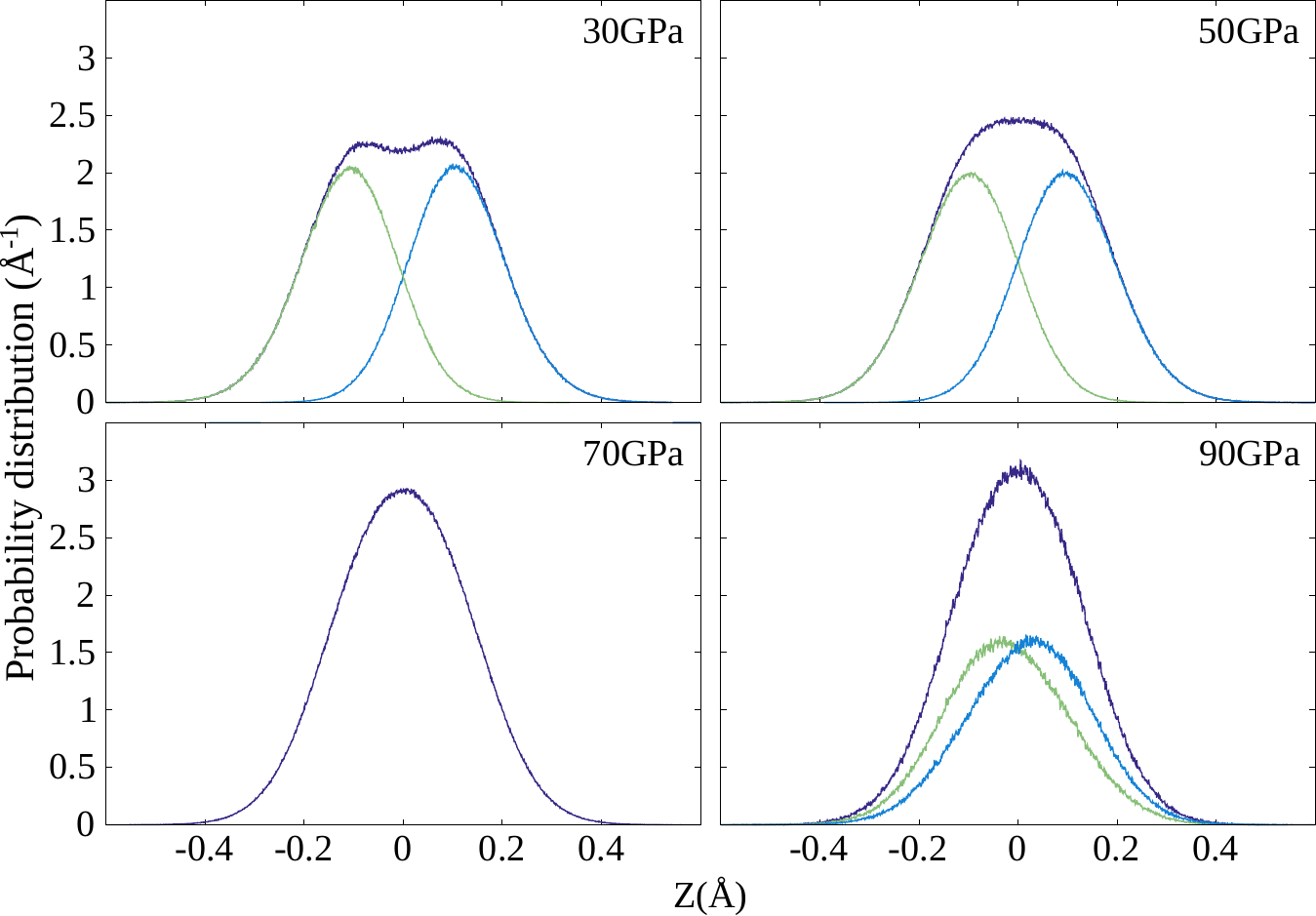}
\caption{Probability distribution of the proton position along $\mathbf{c}$. The green lines relate to the bottom layer of hydrogen nuclei (figure \ref{fig:Description_Brucite}), the blue lines to the top layer, while the purple lines are the sum of both.}
\label{fig:Distribution_z}
\end{figure}

Before addressing proton diffusion, we need to examine the structural changes of the hydrogen layers under increasing pressure and the possible formation of an almost two-dimensional proton layer.
Figure \ref{fig:Distribution_z} shows the probability distribution of protons along the $\mathbf c$ direction. Initially, each proton lies either in the upper or lower plane -- we therefore label the protons according to their initial location and compute the corresponding "lower-layer" or the "upper-layer" proton distributions. 

As pressure increases, the overall distribution width decreases, the double-peak structure still visible at 30 GPa disappears as the lower- and the upper-layer distributions tend to merge (see Fig.~\ref{fig:Distribution_z}). Accounting for NQE via PIMD is crucial  for the study of this process: zero-point energy yields an important contribution to the proton distribution width and makes the merging of the two layers occur at lower pressure than for classical protons.   

An other interesting effect appears when considering hydrogen motion between the two layers. At the lowest pressures (30 and 50 GPa), the bottom and top layer protons do not mix: although proton hopping (either of a few beads or of the whole ring-polymer in the PIMD framework) from one layer to the other does occur, the protons always return to their original layer by a second hop in reverse. Protons initially situated on one layer thus remain on that layer for the whole duration of the simulation with, from time to time, a short exploration of the other layer. However, at 70 GPa, lower and upper layers are not distinguishable as reverse hopping does not always follow: protons do not belong to one particular layer anymore but rather all protons should be considered as forming a single quasi-2D plane within which efficient diffusion can occur, as we further analyze in the following. Interestingly at the highest probed pressure, 90 GPa, the overall distribution becomes narrower, but the lower and upper protons can again be distinguished: as for the lowest two pressures, they tend to stay in their initial layer with only occasional hops, despite a large overlap between the partial distributions. This behavior will be rationalized in the following, examining the different pressure trends of the two hopping mechanisms that are necessary for long-range proton diffusion.

\subsection*{In-plane reorientation}

\begin{figure}[ht!]
\centering
\includegraphics[width=0.7\textwidth]{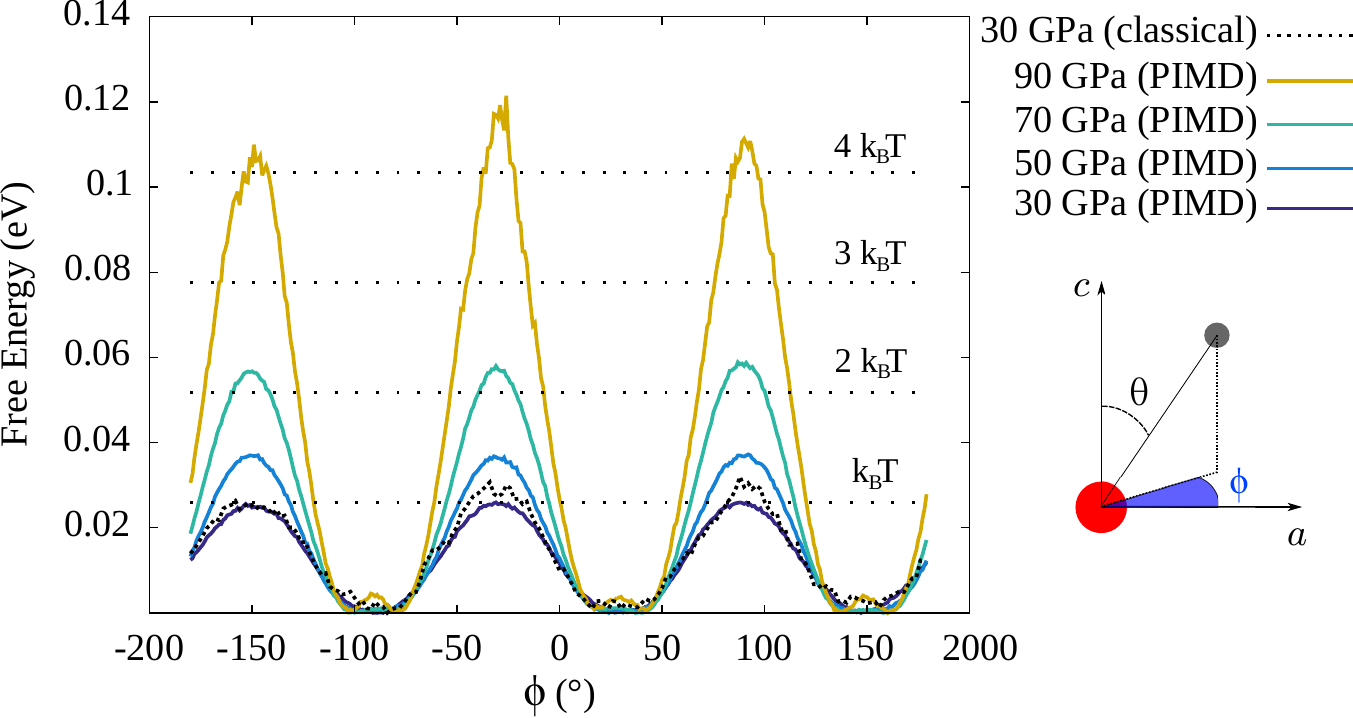}
\caption{Free energy profile along the polar angle $\phi$. 
The proximity between classical and PIMD profiles at 30 GPa shows that the reorientation mechanism is dominated by thermal effects. 
The thermal energies corresponding to $T=300$ K and higher temperatures are also shown as long-dotted lines for comparison.}
\label{fig:Free_E_phi}
\end{figure}

First we discuss the reorientation mechanism. As already described, within the P$\bar{3}$ structure, protons hop in-plane between the $6i$ sites. 
This motion can be efficiently described by the azimuthal angle $\phi$ as shown in the sketch of figure \ref{fig:Free_E_phi}. From the probability distribution of the latter $\mathcal{P}_\mathrm{r}(\phi)$, we extracted the Gibbs free-energy profile $G=-k_B T \log \mathcal{P}_\mathrm{r}(\phi)$, which includes both thermal and quantum effects.

As shown in Figure \ref{fig:Free_E_phi}, the proton free-energy profile along this coordinate has a three-fold symmetry with equivalent barrier heights between the three wells, consistently with the P$\bar{3}$ configuration.
The barrier width of this free energy profile is grossly proportional to $\frac{2\pi}{3}d_\mathrm{O-H}\cos\theta$ ($\theta$ being the zenith polar angle and $d_\mathrm{O-H}$ the covalent O-H bond length), that is, how far the O-H bond slants away from the $\mathbf c$ axis.

Upon compression, we observe that the free energy barrier height increases, from 20 meV at 30 GPa up to 100 meV at 90GPa, revealing a pressure induced confinement of the proton along this coordinate. 
This is in line with the increasing of the average polar angle $\theta$ with pressure that splits the $6i$ sites apart. 
Along with the compression of the layers along the $\mathbf{c}$ axis, the protons from the two layers try to minimize their mutual repulsion by increasing $\theta$.   
As a consequence, the reorientational dynamical disorder, thermally activated at low pressure, tends to slow down.%, eventually to halt.  
We point out that classical simulations, not including NQE, yield almost identical $\mathcal{P}_\mathrm{r}(\phi)$ distributions, meaning that the quantum behavior is in this case moderate within the pressure range that we explored.
The effect of pressure contrasts with that of temperature, which tends to allow the proton to explore equivalently the three wells \cite{dupuis2017quantum} by thermal activation.

\subsection*{Out-of-plane dissociation}

As the hydrogen planes become closer upon compression,
the protons in $6i$ sites adopt quasi-linear O-H-O configurations, where the O anions belong to distinct \ce{Mg(OH)2} stacks. 
This configuration is thus prone to the formation of inter-layer hydrogen bonds.
It has been shown \cite{mookherjee2006high} that only weak hydrogen bonds could be present in brucite. 
However, as we discuss below, a double-well potential is found along the O-O direction at moderate pressure, which suggests the onset of hydrogen bonds and occurs in parallel with  the pressure-induced creation of quasi-2D proton layers in the structure. All along those transformations, NQE play an important role.

\begin{figure}[ht!]
\centering
\includegraphics[width=0.8\textwidth]{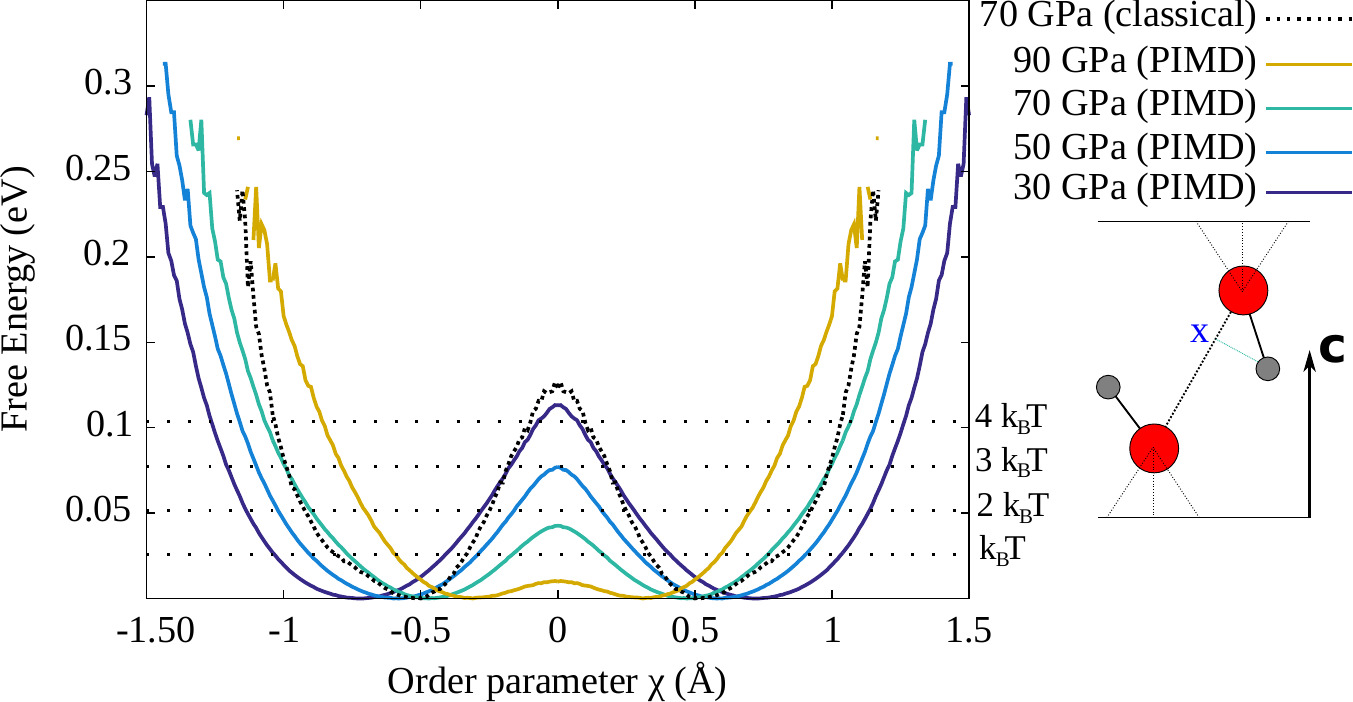}
\caption{Proton free energy profile as a function of the order parameter $\chi$ as defined in the text and sketched on the right (O in red, proton in gray).}
\label{fig:Free_E_chi}
\end{figure}

In order to investigate the proton effective potential, we adopt as the order parameter \cite{Bronstein_2014} $\chi$ which is the difference between the distances that separate the hydrogen atoms from their nearest and second nearest neighbor oxygen atoms projected on the O-O direction (see sketch in figure \ref{fig:Free_E_chi}).
\begin{align}
     \chi = |\overrightarrow{\mathrm{O}_2\mathrm{H}}\cdot \vec{u}_{OO}| - |\overrightarrow{\mathrm{O}_1\mathrm{H}}\cdot \vec{u}_{OO}|
\end{align}
\textrm{with $\vec{u}_{OO}$ the unitary vector in the O$_1$-O$_2$ direction.}

The free-energy profiles along this coordinate (see figure \ref{fig:Free_E_chi}) show that: (i) the effective potential is a double well with minima at $\pm \chi_0$, as for other hydrogen-bond forming crystals \cite{bronstein_AlOOH,Bronstein_2014}; (ii) the barrier height decreases upon compression, from 4 $\mathrm{k_B T}$ at $P=30$ GPa to 0.5 $\mathrm{k_B T}$ at $P=90$ GPa. 
This occurs while the O-O distance shrinks with pressure along the $\mathbf{c}$ axis, bringing the two proton equilibrium positions closer and reducing $|\chi_0|$. 
At high pressures, a proton can therefore hop from one oxygen atom to another through quantum tunneling, zero-point energy or thermal activation, which constitutes the so-called ``dissociation'' process \cite{dupuis2017quantum}.
The difference between classical and quantum simulations is significant as the classical barrier is $\sim 3 \mathrm{k_B T}$ higher than its quantum counterpart at 70 GPa. Therefore, while the proton reorientation mechanism is thermally activated, the dissociation process is mainly quantum driven.

\subsection*{Proton diffusion sweet spot}
\label{sec_sweet_spot}

\begin{figure}[ht!]
\centering
\includegraphics[width=0.7\textwidth]{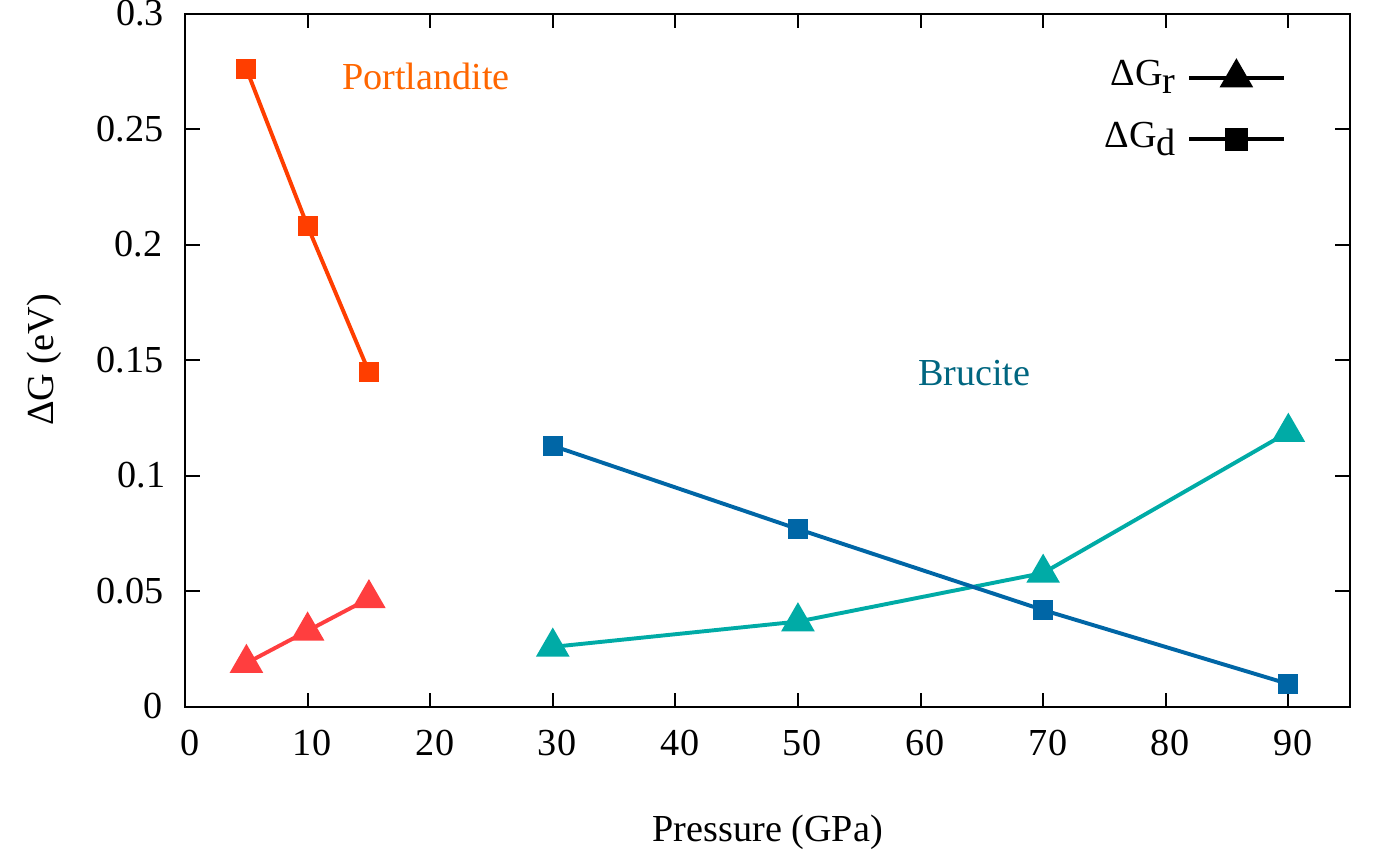}
\caption{Free energy barriers for brucite and portlandite, as computed from the probability distributions for reorientation ($\Delta G_r$) and dissociation ($\Delta G_d$). As discussed in the text, the barriers for proton reorientation and diffusion show opposite trends with increasing pressure, with a non-linear increase of the reorientation barrier in Mg(OH)$_2$. 
The competition between the two trends generates a pressure sweet spot in brucite between 60 and 70 GPa, where both reorientation and dissociation barriers can be overcome; in this pressure range, proton diffusion is enhanced, which requires both mechanisms to occur. Such a sweet spot cannot be reached in portlandite due to the transition taking place at 6 GPa.}
\label{fig:barrier_bru_port}
\end{figure}

The evolution of the free energy barrier heights upon compression is reported in figure \ref{fig:barrier_bru_port}.
In the case of brucite \ce{Mg(OH)2}, the barrier height for dissociation decreases from $\Delta G_d\sim 0.11$ eV at 30 GPa to $\Delta G_d\sim 0.01$ eV at 90 GPa. In contrast, the reorientation barrier increases from $\Delta G_r \sim 0.03$ eV to $\Delta G_r \sim 0.1$ eV within the same pressure range. 
The two curves cross at $\sim$ 70 GPa where hopping rates for the two processes have the same order of magnitude, while reorientation dominates at lower pressures and dissociation prevails at higher pressures. We thus suggest that the highest proton mobility occurs around $P=70$ GPa, which represents the sweet spot for proton diffusion.

We provide in table \ref{tab:reaction_rate} a rough estimate of both dissociation and reorientation reaction rates $\kappa_{d}$ and $\kappa_{r}$, through the Eyring-Polanyi equation \cite{Eyring1935,Polanyi1935}:
\begin{align} \label{eq:EP}
    \kappa_{d,r} = \displaystyle\frac{k_B T}{h}\mathrm{e}^{-\frac{\Delta G_{d,r}}{k_B T}}
\end{align}

While in brucite the typical reorientation and dissociation times are reachable  within our simulation duration, in portlandite (discussed in greater detail below), because of the larger barriers with respect to brucite (see Fig.\ref{fig:barrier_bru_port}), proton dissociation characteristic time in \ce{Ca(OH)2} is on the ns scale or even larger for all pressures before the amorphization transition. We note in passing that $\kappa^{-1}$, as evaluated for classical protons within the Eyring-Polanyi equation, is several order of magnitudes greater.

\begin{table}[ht!]
\centering
\begin{tabular}{|ll|cccc||c|c|c|c|}
\cline{3-10}
\multicolumn{2}{c|}{}& \multicolumn{4}{c||}{Brucite \ce{Mg(OH)2}} &  \multicolumn{4}{c|}{Portlandite \ce{Ca(OH)2}}\\
\hline
$P$ &(GPa) & 30 & 50 & 70 & 90 & 5  &\multirow{3}{*}{\rotatebox{90}{\footnotesize{Transition }}}& 10 & 15  \\
$\kappa_{r}^{-1}$& (ps) & 0.43 & 0.65 & 1.47 & 12.3 & 0.32 & & 0.54 & 0.95 \\
$\kappa_{d}^{-1}$& (ps) & 13.2 & 3.07 & 0.83 & 0.24 & 8.63 $\times 10^3$ & &438 & 44.3 \\
\hline
\end{tabular}
\caption{Eyring-Polanyi inverse reaction rate for the dissociation $\kappa_{d}^{-1}$ and reorientation $\kappa_{r}^{-1}$ mechanisms, computed for brucite (left) and portlandite (right). 
The temperature is $T=300$ K. The transition pressure of 6GPA for portlandite is reported from refs.~\citenum{iizuka_2011,iizuka_2014}}
    \label{tab:reaction_rate}
\end{table}

\subsection*{In-plane proton distribution.}
\begin{figure}[ht!]
\centering
\includegraphics[width=0.85\textwidth]{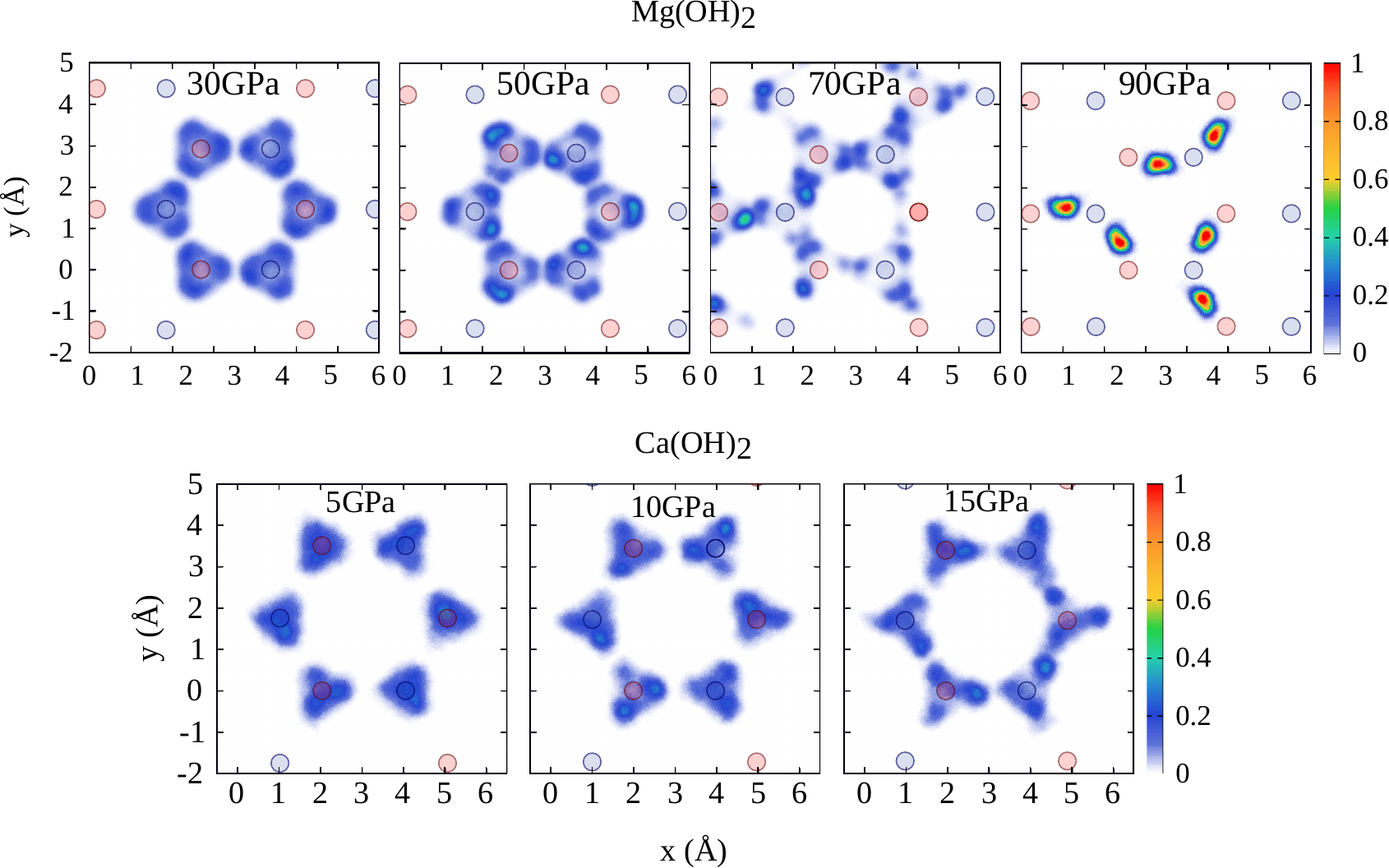}
\caption{Distribution of the proton positions in the $(x,y)$ plane at  30, 50, 70 and 90 GPa for brucite \ce{Mg(OH)2} (upper panels) and portlandite \ce{Ca(OH)2} (lower panels). The circles represent the projection of the oxygen sites on the $(x,y)$ plane (light red: bottom layer, light blue: top layer). Periodic boundary conditions are not used in computing this distribution in order to visualize the displacement of the protons: in brucite, the protons that were initially located within the simulation box move out of its boundaries at 70 GPa, which is an indication of diffusion occurring even within the 30~ps duration of our simulations. 
In portlandite, this might happen at pressures above 15 GPa, thus beyond the structural phase transition pressure reported around 6 GPa from refs.~\citenum{iizuka_2011,iizuka_2014}}.
\label{fig:Distribution_xy_Bru}
\end{figure}

The barrier height analysis above is perfectly consistent
with the probability distribution of the proton positions in the $(\mathbf{a},\mathbf{b})$ plane as shown in figure \ref{fig:Distribution_xy_Bru}. 
For $P=30$ GPa, the proton distribution shows three broad peaks next to each oxygen atom, thus revealing reorientation processes between the $6i$ sites. 
As pressure is increased to 50 GPa, the peaks become narrower and centered at a greater distance from the oxygen sites. This is due to the hindering of the reorientations and the progressive slanting of the O-H bonds towards the $(\mathbf{a},\mathbf{b})$ plane. In addition, the proton density midway between the oxygen sites increases as dissociation become easier. 
At $P=70$ GPa, we observe evidence of a long-range proton diffusion process as the hydrogen nuclei spread all over the simulation box. The onset of dissociation, while reorientation processes still occur, allows the protons to migrate beyond their first neighbors. 
The range of this in-plane motion (over several \AA) can be compared with the width of the out-of-plane $\mathbf c$ axis distribution as in figure \ref{fig:Distribution_z} (approximately 0.2\AA), clearly indicating the two-dimensional character of this process.
Finally, at the highest pressure, $P=90$ GPa, although the dissociation probability keeps increasing, the reorientation processes are locked in. The computed proton distribution is not symmetric among the $6i$ sites anymore and long-range proton diffusion is hindered. 
This confirms the specificity of the 60-70 GPa pressure range as a sweet spot for proton diffusion. 

Due to the relative shortness of the PIMD runs and the fact that a Langevin thermostat with a friction coefficient of 10ps$^{-1}$ is attached to the centroid motion to achieve efficient sampling (see Methods section), the PIMD simulations cannot be used to obtain a direct estimate of the proton diffusion coefficient. Nonetheless, the trends displayed on Fig. \ref{fig:Distribution_xy_Bru}, with the onset of efficient proton diffusion at the sweet spot pressure of 70GPa are qualitatively correct and consistent with the previous discussion. Furthermore, the PILE-L thermostatting scheme\cite{Ceriotti2010_PILE} that we use in this work reduces, in the limit of a vanishing centroid friction, to the Thermostatted Ring-Polymer Molecular Dynamics (TRPMD) algorithm\cite{RPMD_2014} for computing dynamical observables. We expect centroid thermostatting to hinder proton diffusion in our PIMD dynamics, which explains why long-range proton diffusion is not directly observed at 50GPa, even though the estimated dissociation and reorientation inverse rates are relatively short compared to the PIMD trajectory length. Note that centroid thermostatting does not affect the PIMD estimates for static properties such as the free energy barrier $\Delta G$, and therefore it has no impact on the rates presented in Table~\ref{tab:reaction_rate}. Proton diffusion should indeed occur already at pressures lower than 70GPa and it might be at the root of a de-stabilization of the system easing the suggested\cite{hermann2016high} phase transition, or could be involved in the water transfer in the earth mantle.

\subsection*{Comparison with portlandite}

Finally, we close our discussion on proton diffusion in brucite-like minerals by a comparison with portlandite \ce{Ca(OH)2} which has the brucite structure for pressures up to approximately 6 GPa. 
The same analysis as for brucite was done systematically for portlandite. 
In figure \ref{fig:barrier_bru_port} we present the evolution of free energy barriers of the proton reorientation and dissociation mechanisms. We observe that in portlandite the reorientational barrier $\Delta G_r$ at 10 GPa is comparable to that in brucite at 50 GPa. However, the pressure effect on the latter barrier is more important in portlandite as shown by the larger pressure slope of 2.8 meV/GPa while its counterpart in brucite is about 1.6 meV/GPa.
Ca$^{++}$ cations are larger than Mg$^{++}$ ones, so that the computed in-plane O-O distance at $P=10$ GPa is $\simeq$ 3.45 {\AA} in portlandite and $\simeq$ 3.06 {\AA} in brucite; in contrast, the distance between the oxygen anions on distinct stacks is almost the same in the two crystals.  
This implies larger polar angles $\theta$ in portlandite than in brucite, which efficiently hinder the reorientation mechanism, even at relatively low pressures.

As far as the dissociation barriers $\Delta G_d$ are concerned, in portlandite they are greater than in brucite but decrease much faster with pressure. 
The estimated pressure slope $\frac{\partial\Delta G_d}{\partial P}$
is 14 meV/GPa in portlandite, as compared to 2meV/GPa in brucite. 
This derives from the larger compressibility of portlandite with respect to brucite, as demonstrated in a recent work \cite{ulian2019equation}. It has to be noticed that the large value of the dissociation barrier in portlandite would require very long runs of path integral ab initio molecular dynamics, much beyond the scope of this work, to record a significant number of events. Nevertheless, we detected a few events as some of the replicas of the PIMD simulations did occasionally reach the barrier top. A precise evaluation of the barrier heights would have nevertheless required the use of accelerated sampling techniques, such as in ref.~\citeonline{Ivanov2015}.

Finally, the crossing point of the dissociation and orientational barriers in portlandite should occur beyond 20 GPa, with diffusion barriers comparable to that of brucite at 70 GPa. However, a transition towards another phase is reported at 6GPa \cite{iizuka_2011,iizuka_2014} and our own simulations reveal instability of the system at 20 GPa. Therefore, as shown in figure \ref{fig:Distribution_xy_Bru}, no diffusion was observed for portlandite within the time scale of our simulations. Indeed, the reaction rate estimates, given in table \ref{tab:reaction_rate}, yield much longer times than in brucite.

The values reported in Table \ref{tab:reaction_rate} for portlandite agree remarkably well with the results of Dupuis and coworkers\cite{dupuis2017quantum} as regards the reorientation rate at 10 GPa and its increase with decreasing pressure. Our results for the dissociation rate, however, are lower than that of ref.~\citeonline{dupuis2017quantum} by as much as 3 orders of magnitude under comparable thermodynamic conditions. In ref.~\citeonline{dupuis2017quantum}, different methodologies were used for the computation of the two rates: whereas path integral methods were employed for the  reorientation process, metadynamics with classical nuclei was used for dissociation. In this work on the other hand, we compute the two rates within a fully consistent framework: the same PIMD trajectories are employed to estimate the free energy barriers associated with the two processes, which are then inserted into the Eyring-Polanyi approximation. We do not resort to any type of accelerated sampling scheme to overcome the barriers. Note that for the lowest pressure of 5 GPa, the ring-polymer beads very rarely explore the top of the dissociation barrier, therefore the uncertainty on $\Delta G_d$ is large (we estimate $\Delta G_d = \pm$ 20 meV). This accuracy is nonetheless sufficient to exclude, within our approach, such a fast dissociation as reported in ref.~\citeonline{dupuis2017quantum}.

It can be noted, that among the other systems sharing the same structure as brucite, our preliminary analysis of theophrastite \ce{Ni(OH)2} indicates that this system should also present a sweet spot for proton diffusion at approximately the same pressure as \ce{Mg(OH)2}, due to comparable ionic radii between Mg$^{++}$ and Ni$^{++}$ cations.
 
\section*{Conclusion}
To summarize, we analyzed the proton diffusion mechanism in both brucite (\ce{Mg(OH)2}) and portlandite (\ce{Ca(OH)2}) under pressure, taking into account nuclear quantum effects by path-integral based ab initio molecular dynamics. 
Proton diffusion in those crystals involves two stages to occur: a reorientation motion within the ($\mathbf{a}$,$\mathbf{b}$) plane, and a proton dissociation between two oxygen atoms on opposite layers. 
Firstly, we have seen that the reorientation mechanism is thermally activated and that the pressure tends to localize the proton in a certain orientation, making the reorientation motion less likely. 
Secondly, in contrast with the reorientation, we showed that the dissociation mechanism was quantum driven and was made easier by increasing pressure through the formation of a quasi two-dimensional hydrogen layer.

These two antagonistic effects give rise to a pressure sweet spot for proton diffusion within 60 and 70 GPa in brucite. However, proton diffusion could also occur at much lower pressure, although it is less probable, and could be at the root of a destabilization of the structure, as suggested by the theoretical predictions of a phase transition \cite{hermann2016high} at 20 GPa or decomposition into \ce{MgO} and \ce{H2O} at the same pressure\cite{fei1993static}. 
Beyond this pressure threshold the reorientation becomes a bottleneck for proton diffusion, while dissociation is the rate-limiting step at lower pressure.

Finally, by systematic comparison with portlandite, we demonstrate the specificity of brucite for proton diffusion. Indeed, the proton diffusion sweet spot in portlandite would occur at pressures well beyond its transition toward an another phase.

\section*{Methods}

Molecular dynamics (MD) simulations were carried out at room temperature and fixed volume either via a classical Langevin equation or within the Path Integral (PI) framework to take into account nuclear quantum effects while the electronic structure and atomic forces were described within the Density Functional Theory (DFT).
We compute the electron density and atomic forces within the Perdew-Wang Generalized Gradient approximation to the exchange and correlation density functional \cite{PW91}, using the Quantum Espresso package \cite{giannozzi2009quantum} in combination with the i-PI\cite{Ceriotti14} interface for the path-integral simulations.
We employed ultra-soft pseudo-potentials with a plane wave expansion cutoff of $E_{cut} = 50$ Ry for the Kohn-Sham states and 8 times as large for the charge and the potential, ensuring total energy convergence.
Both brucite \ce{Mg(OH)2} and portlandite \ce{Ca(OH)2} were simulated by using a hexagonal $(\sqrt{3} \times \sqrt{3} \times 1)$ supercell and a $(2 \times 2 \times 2)$ k-point sampling centred at $\frac{2\pi}{a}( \frac{1}{2}, \frac{1}{2}, \frac{a}{2c})$.

The number of beads in the PIMD simulations was set to 24 and checked to provide convergence of kinetic and potential energies. The PIMD simulations were performed using the efficient PILE-L thermostatting scheme\cite{Ceriotti2010_PILE} with a centroid friction coefficient of 10 ps$^{-1}$.
Lattice parameters were obtained through systematic volume relaxation of the system ensuring isotropic stress tensors for each pressure.
The optimized equilibrium lattice parameters $a$ and $c$ were remarkably similar between classical and path-integral MD simulations for the pressures here considered: the differences between the classical and quantum results for $a$ and $c$ amounted to few hundredths of {\AA}.
Finally, the typical duration time of the simulations was 30 ps.

\bibliography{Bibliography}

\section*{Acknowledgements}
This work was granted access to the HPC resources of CINES under the allocation A0030906719 made by GENCI.

\section*{Author contributions statement}
S.S., P.D., and F.F. designed research; S.S. carried out the simulations; S.S., P.D., and S.H. analyzed the results; S.S. and P.D. wrote the manuscript that was jointly revised by all of the authors. 

\section*{Additional information}
%To include, in this order: \textbf{Accession codes} (where applicable)
\textbf{Competing interests}. 
The authors declare no competing interests.

\end{document}